%% file: root.tex
\documentclass[letterpaper, 10 pt, conference]{ieeeconf}  
\usepackage{graphicx}                                             
\usepackage{subcaption} 
\usepackage{array} 
\usepackage{adjustbox}   
\usepackage{mwe} 
\usepackage{amsmath} 
\usepackage{amssymb} 
\usepackage{tikz}
\usepackage{color}
\definecolor{darkgreen}{rgb}{0,0.5,0}
\usepackage[colorinlistoftodos]{todonotes}
\usepackage{soul}
\sethlcolor{yellow} 
\input{macros}

\IEEEoverridecommandlockouts
\title{\LARGE\bf
Energy-Aware Lane Planning for Connected Electric Vehicles in Urban Traffic: Design and Vehicle-in-the-Loop Validation
}

\author{Hansung Kim\textsuperscript{1}, Eric Yongkeun Choi\textsuperscript{1}, Eunhyek Joa\textsuperscript{3}, Hotae Lee\textsuperscript{1}, Linda Lim\textsuperscript{2}, \\ Scott Moura\textsuperscript{2}, and Francesco Borrelli\textsuperscript{1} \thanks{HK, EYC, HL, and FB are with the Department of Mechanical Engineering, UC Berkeley. LL and SM are with the Department of Civil Engineering, UC Berkeley. EJ is with Zoox, Inc. Email: hansung@berkeley.edu}}

\begin{document}
\maketitle
\thispagestyle{empty}
\pagestyle{empty}

\begin{abstract}
Urban driving with connected and automated vehicles (CAVs) offers potential for energy savings, yet most eco-driving strategies focus solely on longitudinal speed control within a single lane. This neglects the significant impact of lateral decisions, such as lane changes, on overall energy efficiency—especially in environments with traffic signals and heterogeneous traffic flow. To address this gap, we propose a novel energy-aware motion planning framework that jointly optimizes longitudinal speed and lateral lane-change decisions using vehicle-to-infrastructure (V2I) communication. Our approach estimates long-term energy costs using a graph-based approximation and solves short-horizon optimal control problems under traffic constraints. Using a data-driven energy model calibrated to an actual battery electric vehicle, we demonstrate with vehicle-in-the-loop experiments that our method reduces motion energy consumption by up to 24\% compared to a human driver, highlighting the potential of connectivity-enabled planning for sustainable urban autonomy.
\end{abstract}

\section{Introduction}

Connected and Automated Vehicles (CAVs) can improve safety, traffic flow, 
and energy efficiency by coordinating with traffic signals and nearby vehicles 
through vehicle-to-infrastructure (V2I) and vehicle-to-vehicle (V2V) communication \cite{GUANETTI201818}. 
Eco-driving strategies leveraging Signal Phase and Timing (SPaT) data from connected 
traffic lights have demonstrated 11--16\% energy savings \cite{Levin2023, Rousseau}, 
but these methods typically assume fixed-lane driving and optimize only longitudinal motion \cite{farias, sangjae, joa_ioniq5}. 
In practice, lane changes also influence efficiency by avoiding slower vehicles 
or aligning with favorable traffic signals \cite{sangjae}. 
Yet most eco-driving studies overlook this factor, limiting their applicability 
to realistic multi-lane scenarios. 
We extend eco-driving to include lane-change decisions for energy savings for CAVs. We focus on how connectivity (V2I/V2V) can help an energy-aware lane selection strategy, extending conventional eco-driving to a multi-lane context.

In this work, we propose a novel motion planning framework that integrates lane-change decisions with energy-efficient speed optimization, enabled by V2I/V2V communication. Our approach uses a hierarchical control structure: a high-level decision module selects the most energy-efficient lane (leveraging SPaT data and V2V traffic information), and a low-level trajectory planner computes a safe and smooth trajectory to that lane. We validate the proposed system in a vehicle-in-the-loop (VIL) testing environment, embedding a real CAV into a virtual traffic simulation. This setup provides a real evaluation of energy savings and driving performance with physical vehicle dynamics and realistic surrounding traffic, bridging the gap between pure simulation and real-world public road deployment.
 As with all literature on this topic, the impact of deploying this technology at scale on global energy improvement is not studied in this work.

\section{RELATED WORKS}
\subsection{Eco-Driving with V2I Connectivity}
A large body of research has explored energy-efficient driving using V2I communication, particularly along signalized corridors. Optimization-based eco-driving controllers have been proposed to leverage SPaT data over corridors with multiple traffic lights. For example, in \cite{ard2023VILCAV}, the authors apply Pontryagin’s Minimum Principle to develop an eco-driving controller, demonstrating fuel savings of up to 36\% compared to a human-modeled driver through VIL experiments. In \cite{bae2022ecological}, the authors design a real-time eco-driving controller formulated as a receding-horizon optimal control problem and demonstrate its effectiveness through real-world road testing, achieving a 31\% energy reduction at the cost of increased travel time. These works demonstrate that CAVs can significantly improve efficiency through longitudinal speed planning, when leveraging infrastructure information. However, these studies assume a fixed lane. The controllers react to surrounding traffics and red lights, but do not consider proactive lane changes to avoid deceleration.

\subsection{Lane Changes in Eco-Driving}
Only recently have researchers begun to incorporate lateral maneuvers into eco-driving. 
In \cite{dong2024lc}, the authors note that most studies emphasize longitudinal control and neglect 
lane changes, which can reduce efficiency when following slower vehicles. To address this, 
they propose a graph-search strategy that selects lane sequences based on predicted traffic, 
though the effect of traffic lights is not considered. Similarly, \cite{yang2023lc} 
designs an overtaking strategy in which an energy-optimal speed trajectory is first computed, 
and a lane change is triggered if a slower vehicle disrupts that plan. These studies indicate 
that lane-changing can enhance longitudinal eco-driving, yet their validation remains limited 
to simulation. Few works have demonstrated lane-changing eco-driving in real vehicles or 
high-fidelity VIL platforms, aside from our previous work \cite{arpae-lc}.

\section{Contributions} This work's contributions to energy-efficient motion planning for connected, automated urban driving are as follows

\begin{itemize} 
    \item Formulation of the lane selection problem in urban 
    environments as a joint behavior planning task over 
    lateral and longitudinal actions, using SPaT and 
    surrounding traffic information. 
    \item A hybrid lane selection algorithm that combines 
    short-horizon optimal control with a graph-based 
    approximation of long-term energy costs, supported by 
    a data-driven energy model calibrated on a Hyundai IONIQ5.  
    \item Vehicle-in-the-loop experiments demonstrating up to 
    24\% motion energy reduction compared to human driving 
    and 7\% savings over a lane-keeping eco-driving baseline.
    \item Highlights the role of eco-driving technologies in electric vehicles, where regenerative braking diminishes the benefit of traditional stop-and-go minimization. This emphasizes the need for a refined understanding of energy-optimal driving behavior for electric vehicles.
\end{itemize}
In summary, this work presents a modular framework for energy-aware lane selection in connected electric vehicles and provides experimental insights that support its potential for improving energy efficiency in urban driving scenarios.
Understanding average savings on daily commutes is complex and outside the scope of this work.

\section{Problem Setup}
\begin{figure}
    \centering
    \includegraphics[width=0.75\columnwidth]{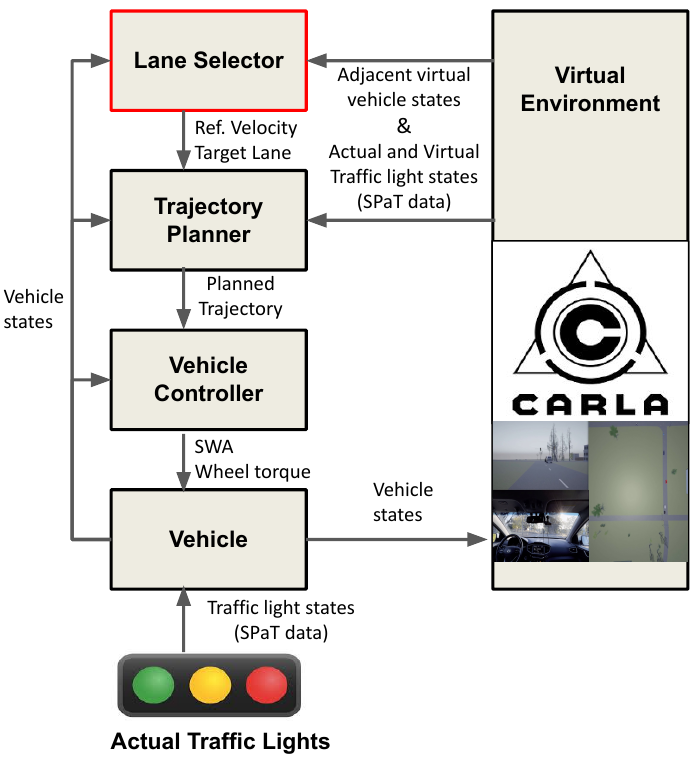}
    \caption{Vehicle-in-the-loop control architecture \cite{arpae-lc}}
    \label{fig:system_arch}
    \vspace{-0.5cm}
\end{figure}

In this section, we formulate the problem of energy-efficient lane selection for connected, automated urban driving. We adopt the hierarchical control architecture introduced in \cite{arpae-lc} and focus on designing an energy-efficient lane selector for battery electric vehicles (BEVs) in urban environments. The architecture, illustrated in Fig. \ref{fig:system_arch}, and the corresponding vehicle-in-the-loop (VIL) setup are detailed in \cite{arpae-lc}. 
We assume that a route from point A to point B is given and that the traffic light cycles are deterministic.
\subsection{Lane Selection Problem}
\begin{figure}
    \centering
    \includegraphics[width=0.8\columnwidth]{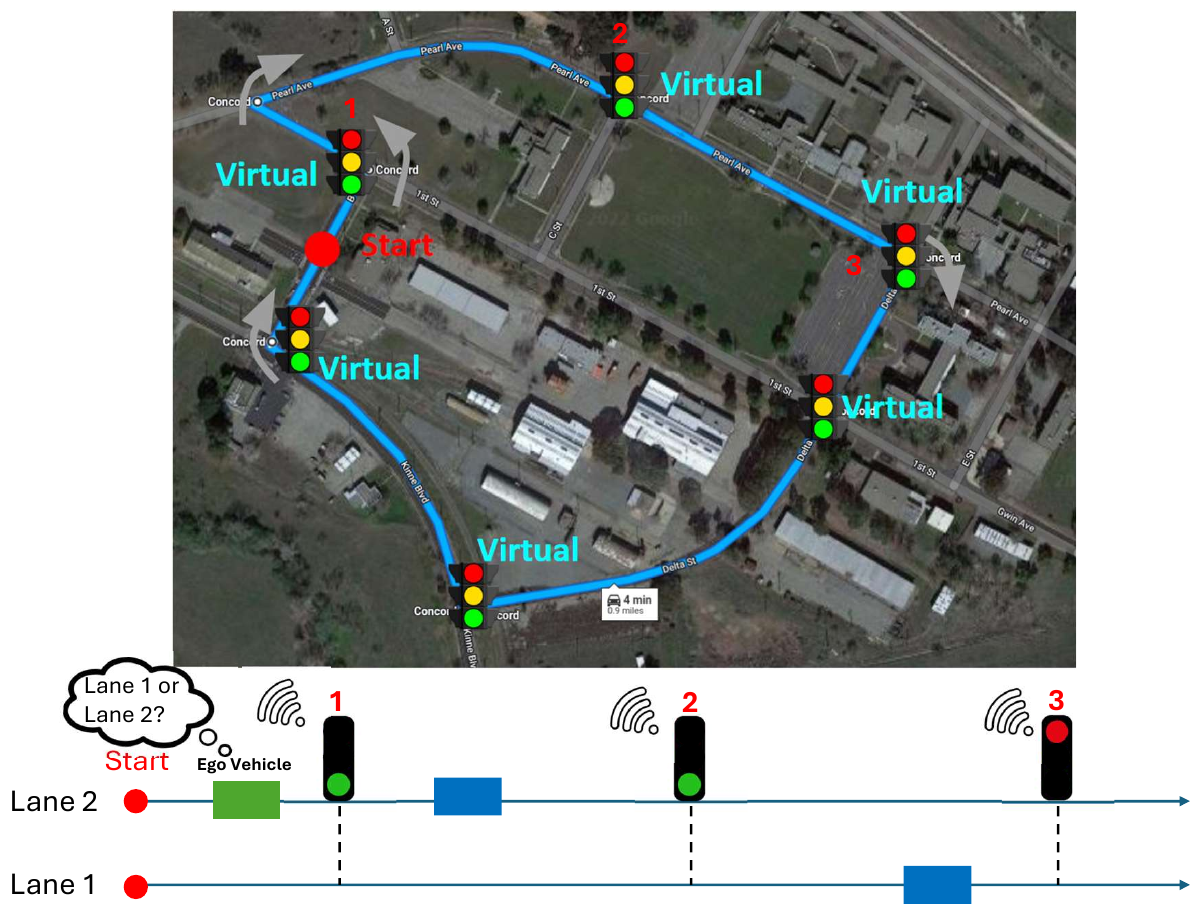}
    \caption{Lane selection problem in urban road}
    \label{fig:problem_statement}
    \vspace{-0.5cm}
\end{figure}
This paper focuses on a two-lane, one-way road, as illustrated in Fig. \ref{fig:problem_statement}, although the proposed methods can be extended to roads with more than two lanes. We assume that the CAV can perfectly observe surrounding vehicles' states within a certain radius and, by leveraging V2I connectivity, can receive SPaT information for the next $N_{TL}$ traffic lights along its route. The information available to the lane selector includes the controlled vehicle's state, denoted as $x^{EV}(t)=[s(t),v(t)]^\top$, where $s(t)$ and $v(t)$ are the longitudinal displacement along the centerline of the given route and the longitudinal velocity, respectively. Additionally, the vehicle's lane information is encoded as a categorical variable $lane(t)\in \{0, 1\}$. The surrounding vehicles (SV) within the detection range are collected into 
\begin{align}\label{eq:NPC}
o(t)=[x^i(t),lane^i(t)]_{i=1}^{N_{SV}}, 
\end{align}
where $N_{SV}$ is the number of detected SVs.
Lastly, the SPaT of the next $N_{TL}$ upcoming traffic lights is collected into 
\begin{align}\label{eq:SPaT}
SPaT(t)=\{s_{tl}^i,p_{tl}^i(t), t_{tl}^i(t)\}_{i=1}^{N_{TL}},
\end{align}
where $s_{tl}^i\in \mathbb{R}$, $p_{tl}^i(t)\in \{r,g,y\}$, $t_{tl}^i(t) \in \mathbb{R}$ denote the longitudinal position of the $i$-th traffic light along the centerline of the route, the current signal phase (red, green, or yellow), and the remaining time in the current phase, respectively. The duration of the full cycle of traffic signals—green ($t^i_g$), yellow ($t^i_y$), and red ($t^i_r$)—are assumed to be estimated from historic SPaT data. Let $\mathcal{C}_{\text{pass}}$ denote the lane selector's decision regarding whether the controlled vehicle will pass the upcoming traffic light and which lane it should proceed in. In the case of two lanes, 
\begin{equation}\label{eq:C_pass}
\mathcal{C}_{\text{\scriptsize pass}} \in \mathcal{LS} := \{\text{\scriptsize PASS0}, \text{\scriptsize PASS1}, \text{\scriptsize NONPASS0}, \text{\scriptsize NONPASS1}\},
\end{equation}
where PASS0 and PASS1 denote the lane selector's decision to pass the upcoming traffic light in lane 0 and lane 1, respectively. NONPASS0 and NONPASS1 denote the decision to not pass the upcoming traffic light in lane 0 and lane 1, respectively.
The subset of passing decisions is denoted by $\mathcal{LS}_{\text{PASS}} := \{\text{PASS0}, \text{PASS1}\}$, and the subset of nonpassing decisions is denoted by $\mathcal{LS}_{\text{NONPASS}} := \{\text{NONPASS0}, \text{NONPASS1}\}$.

Now, the lane selection problem is to design a mapping function 
\vspace{-0.4cm}
\begin{align}\label{eq:laneselection_map}
\small \mathcal{C}_{\text{pass}, t}=f(x^{\text{EV}}(t),o(t),SPaT(t)),
\end{align}
where $\mathcal{C}_{\text{pass}, t}$ is the lane selection decision at time $t$.

\subsection{Energy-efficient Lane Selection}
For energy-efficient lane selection, it is crucial to understand how the lane selector's current decision influences the vehicle's long-horizon energy consumption, in the context of current traffic information. To achieve this, we first develop a model for the vehicle's energy consumption.

\subsubsection{Data-driven Energy Consumption Model \cite{joa_ioniq5}}
We approximate the energy consumption of the battery electric vehicle (BEV) given a longitudinal trajectory. Specifically, we assume that the energy consumption is a function of the longitudinal displacement along the centerline and the longitudinal acceleration of the vehicle. For more details, refer to Sec. II.C. in \cite{joa_ioniq5}. By recording the vehicle's state-of-charge (SOC) and longitudinal motion, we fit a quadratic regression model to estimate the energy consumption rate $\Delta E$, denoted as $\mathcal{E}(v, a)$, where $v$ and $a$ are the vehicle's longitudinal velocity and acceleration. The quadratic regression model is parameterized as 
\begin{equation} \label{eq:energy}
    \small \mathcal{E}(v, a)= z^\top P z + q^\top z + r
\end{equation}
where $z = [v, a]$. The parameters $P \in \mathbb{R}^{2 \times 2}$, $q \in \mathbb{R}^2$, and $r \in \mathbb{R}$ are constants, with $P \succ 0$, making \eqref{eq:energy} a convex quadratic function. 

\begin{figure}
    \centering
    \includegraphics[width=0.8\columnwidth]{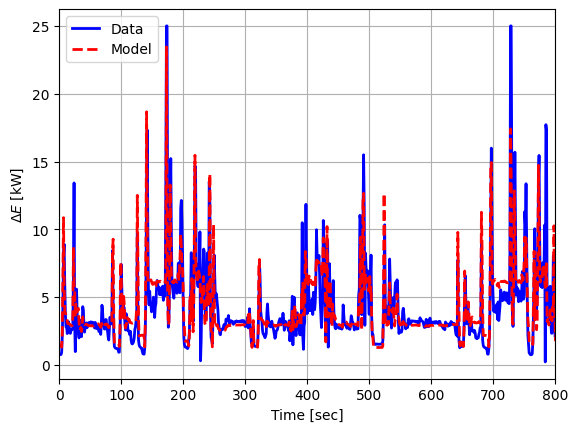}
    \caption{Comparison of quadratic energy consumption model and measured data}
    \label{fig:energy_model}
    \vspace{-0.5cm}
\end{figure}

The resulting regression and the corresponding measured energy consumption data are shown in Fig.~\ref{fig:energy_model}. The mean error is approximately 1\% \cite{joa2024ioniq5journal}. The observed discrepancy between the model and the measured data is attributed to unmodeled non-linear effects such as aerodynamic drag, rolling resistance, and variations in motor and battery efficiency. Note that this approximate energy consumption model is  used to reason about energy consumption for lane selection, but not used to measure the actual energy consumption of the vehicle. We directly read and record the SOC from the vehicle's CAN signals.
Given the energy consumption model of the vehicle, we can reason about long-horizon energy-efficiency in designing an energy-efficient lane selector: 
\begin{align} \label{eq:eco_ls}
\small{\mathcal{C}^{\text{eco}}_{\text{pass},t}=f(x^{\text{EV}}(t),o(t),SPaT(t),\mathcal{E}).}
\end{align}

\section{Proposed Lane Selection Methods}
In this section, we outline the proposed method for energy-efficient lane selection leveraging V2I connectivity. Let the globally energy optimal lane selection strategy for following a given route be  $\{\mathcal{C}^{\star,\text{eco}}_{\text{pass},t},\mathcal{C}^{\star,\text{eco}}_{\text{pass},t+1},...,\mathcal{C}^{\star,\text{eco}}_{\text{pass},t+T}\}$, where $T$ is the task horizon. In general, solving for this globally energy optimal lane change strategy for a large $T$ is intractable due to imperfect forecasts of the NPC vehicles and vehicle dynamics modeling errors. Thus, we approximate it by solving for a locally energy optimal lane selection strategy in a receding horizon fashion. Every $\Delta t_{LS}$, we compute a new locally energy optimal lane change strategy and reference velocity profile for the trajectory planner.

The proposed method is a hybrid approach that solves $|\mathcal{LS}|$ short finite-horizon optimal control problems up to the upcoming traffic light (i.e. one for each pass/nonpass and lane choice combination). The energy cost beyond the upcoming traffic light—associated with passing or not passing the next $N_{TL}$ traffic lights and selecting the appropriate lane—is then approximated using a graph search method and incorporated as a terminal cost-to-go. Finally, the lane selection decision with the minimum total cost is chosen.

\subsection{Optimization and Graph Search Method}
In this method, we formulate a finite-horizon optimal control problem that considers the signal and phase of the next upcoming traffic light, as well as the safety constraints with respect to other NPC vehicles, based on \(\mathcal{C}_{pass}\).

\subsubsection{Vehicle Model}
We model the longitudinal dynamics of the vehicle as a double integrator: 
\begin{align} \label{eq:vehicle_model}
\dot{x}(t) = \begin{bmatrix}
    0       & 1  \\
    0       & 0  \\
\end{bmatrix} x(t) + \begin{bmatrix}
    0  \\
    1   \\
\end{bmatrix} u(t),
\end{align}
where $u(t)=a(t)$ is the longitudinal acceleration of the vehicle. We discretize the dynamics \eqref{eq:vehicle_model} using the sampling time $\Delta t$ of 0.5 seconds. The discretized dynamics are denoted as 
\begin{align} \label{eq:vehicle_model_dt}
x_{k+1}= Ax_k + B u_k,
\end{align}
where $x_k$ and $u_k$ are state and control input at time step $k$, respectively.
\subsubsection{Graph Search Terminal Cost Approximation}
\begin{figure}
    \centering
    \includegraphics[width=\columnwidth]{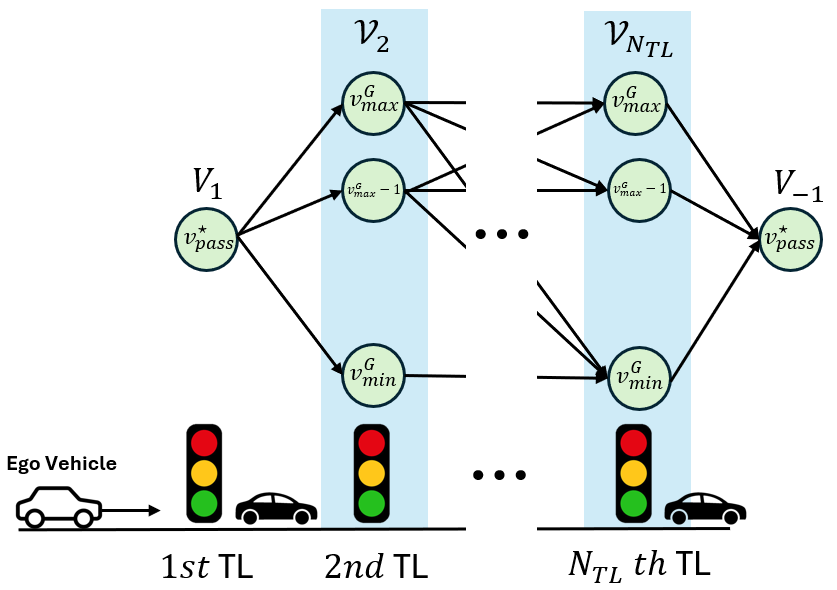}
    \caption{Illustration of a graph for navigation through multiple traffic lights. The graph approximates the pass or nonpass decision for a fixed lane and velocity}
    \label{fig:graph}
    \vspace{-0.5cm}
\end{figure}

We simplify the lane selection problem for passing the $N_{TL}$ traffic lights by assuming the vehicle travels in a fixed lane and at a constant velocity between each traffic light (nodes) and formulating it as a directed graph defined as
\begin{align} \label{eq:graph}
    G=(V,E),
\end{align}
where $V = \{V_1, \mathcal{V}_2, \dots, \mathcal{V}_{N_{TL}}, V_{-1}\}$ and $\mathcal{V}_i$, $\forall i\in\mathbb{I}_{1}^{N_{TL}}$, represents the subset of nodes corresponding to each traffic light as depicted in Fig. \ref{fig:graph}. Each node in $\mathcal{V}_i$ represents the velocity at the point of passing the $i$-th traffic light. We grid the passing velocities from $v^{G}_{min} = 5\,\text{m/s}$ to $v^{G}_{max} = 10\,\text{m/s}$ with a grid size of $1\,\text{m/s}$. The starting and terminal nodes $V_1$ and $V_{-1}$ represent the first traffic light ahead and the ego vehicle's position after passing the $N_{TL}$-th traffic light, respectively. The solution of the optimal control problem determines the passing velocity at the 1st traffic light. For simplicity, the passing velocity at the terminal node is set equal to that of the starting node.

The edges represent the cost of stopping at a traffic light—specifically, the vehicle must stop at the $j$-th traffic light if the signal is not green at the time it arrives at the traffic light ($t_j$), i.e., if $p_{tl}^{j}(t_j) \neq g$—as well as the approximate change in kinetic energy due to velocity transitions between nodes. The edge cost for the directed connection $V_i \rightarrow V_j$ is computed as follows:
\begin{align}
c(V_i,V_j) = v(V_i)^2 - v(V_j)^2 -\mathbb{1}_{\{p_{tl}^{j}(t_j) \neq g\}}v(V_j)^2,
\end{align}
where $v(V)$ represents the passing velocity value of the node and $\ell$ denotes the $\ell$-th traffic light that the node $V_j$ represents. $tl^{\ell,p}$ at the time the ego vehicle reaches the traffic light is computed based on the travel time between nodes assuming the vehicle travels at a constant velocity between traffic lights. $\mathbb{1}(\cdot)$ is the Heaviside step function. The connections from nodes in $\mathcal{V}_{N_{TL}}$ to the terminal node $V_{-1}$ are added to penalize the velocity deviation from the velocity at the point of passing the first traffic light. Alternatively, the data-driven energy model in \eqref{eq:energy} can be used to approximate the energy consumption resulting from changes in node velocities, assuming constant acceleration. However, for simplicity, this work adopts the kinetic energy model to compute edge costs. Note that the graph \eqref{eq:graph} satisfies additive property meaning $c(V_a,V_c) = c(V_a,V_b) + c(V_b,V_c)$
for any node connections $V_a\rightarrow V_b$ and $V_b \rightarrow V_c$. Additivity allows us to exploit dynamic programming and optimal substructure, formulating the problem as a path-finding problem and applying Dijkstra's algorithm \cite{dijkstra1959note} to compute $\min c(V_1, V_{-1})$ given $v(V_1)$. We define the minimum sum of the edge cost from the starting node to the terminal node as
\begin{align}
    J^\star_{graph}(v(V_1)) = \min_{\{V_i \in \mathcal{V}_i\}_{i=2}^{N_{TL}} } c(V_1, V_{-1}), \label{eq:J_graph}
\end{align}
and use it to approximate the energy cost from the 1st traffic light to the $N_{TL}$-th traffic light, assuming the ego vehicle travels in the fixed lane and at a constant velocity.
\subsubsection{Constraints}
The ego vehicle is subject to state and input constraints defined below
\begin{align}
\mathcal{X}&:=\{x\in\mathbb{R}^2|\ 2 \leq v \leq 11, \text{where} \; x=[s,v]^\top \}, \\
    \mathcal{U}&:=\{u\in\mathbb{R}|-4 \leq u \leq 2\}.
\end{align}
Collision avoidance constraints with respect to the NPC vehicles are also considered, and their set representation is denoted as $\mathcal{F}^{LK}(s^{\text{front}}, v^{\text{front}})$, where \(s^{\text{front}}\) and \(v^{\text{front}}\) denote the states of the vehicle immediately preceding the ego vehicle in the same lane. We assume that the NPC vehicles maintain constant velocity during the prediction horizon \(N\). For more details, refer to \cite{arpae-lc}.

Additionally, conditional on the next traffic light's current signal phase at time, $p_{tl}^{\text{next}}(t)$, the ego vehicle is subject to traffic light passing or stopping constraints as follows
\begin{align}
    \mathcal{P}(s_{tl}^{\text{next}})=\{x\in \mathbb{R}^2 | s \geq s_{tl}^{\text{next}} + d\}, \\
    \mathcal{S}(s_{tl}^{\text{next}})=\{x\in \mathbb{R}^2 | s_{tl}^{\text{next}}-d \geq s\},
\end{align}
where $s_{tl}^{\text{next}}$ is the next traffic light's relative longitudinal position along the centerline of the given route and $d$ is the user-defined margin for controlling the conservativeness of the lane selection planner. In this work, we use $d=3$ m. Further, these traffic light passing and stopping constraints are time-varying constraints, which means the time steps that these constraints are applied depend on the pass or nonpass decision $\mathcal{C}_{\text{pass}}$ and current phase of the next traffic light $p_{tl}^{\text{next}}(t)$. The lower and upper bounds on the time for pass and stop constraints are as follows
\begin{subequations}
  \begin{align}
    \bar{t}_{\text{pass}}(\mathcal{C}_{\text{pass}},p_{tl}^{\text{next}}(t)) &= CT(\phi,\mathcal{C}_{\text{pass}})- t_g^{\text{margin}},\\
    \underline{t}_{\text{pass}}(\mathcal{C}_{\text{pass}},p_{tl}^{\text{next}}(t))&=CT(\phi,\mathcal{C}_{\text{pass}}) + t_g^{\text{margin}},\\
    \bar{t}_{\text{stop}}(\mathcal{C}_{\text{pass}},p_{tl}^{\text{next}}(t)) &=CT(\phi,\mathcal{C}_{\text{pass}}),\\
    \underline{t}_{\text{stop}}(\mathcal{C}_{\text{pass}},p_{tl}^{\text{next}}(t)) &= 0,
\end{align}  
\end{subequations}
where $\bar{t}_{\text{pass}}$ and $\underline{t}_{\text{pass}}$ denote the upper and lower bounds of the time interval during which the pass constraint is active; analogous definitions apply to the stop constraint. The parameter $t_g^{\text{margin}}$ is a user-defined safety margin that ensures the vehicle passes through the traffic light within the green phase. In this work, we set $t_g^{\text{margin}} = 2$ s. Additionally, the function $CT$ denotes the cycle time adjustment, which accounts for the current signal phase of the upcoming traffic light and the decision to either pass or stop. Let $\phi \in \{0,1,2\}$ be the categorical encoding of the next traffic light's signal phase, corresponding to green, yellow, and red respectively: $[0 \mapsto g,\ 1 \mapsto y,\ 2 \mapsto r]$. The next traffic light's cycle durations is $\text{Cycle} := [t^{\text{next}}_g, t^{\text{next}}_y, t^{\text{next}}_r],$
where $t^{\text{next}}_g$, $t^{\text{next}}_y$, and $t^{\text{next}}_r$ denote the durations of the green, yellow, and red phases respectively—which can be obtained from the traffic light's SPaT data. The $CT$ function is defined as follows:
\begin{align}
&CT(\phi, \mathcal{C}_{\text{pass}}) := t_{tl}^{\text{next}}(t) +\hspace{-0.6em} \sum_{i=1}^{\delta(\phi, \mathcal{C}_{\text{pass}})} \hspace{-0.6em} \text{Cycle}[(\phi + i) \bmod 3],
 \\
& \delta(\phi, \mathcal{C}_{\text{pass}}) = 
\begin{cases}
0 & \text{if } \mathcal{C}_{\small{\text{pass}}} \in \mathcal{LS}_{\text{PASS}} \\
2 & \text{if } \mathcal{C}_{\small{\text{pass}}} \in \mathcal{LS}_{\text{NONPASS}}  \wedge \phi = 2\ (\small{\text{r}}) \\
1 & \text{if } \mathcal{C}_{\small{\text{pass}}} \in \mathcal{LS}_{\text{NONPASS}} \wedge \phi = 0\ (\small{\text{g}}) \\
2 & \text{if } \mathcal{C}_{\small{\text{pass}}} \in \mathcal{LS}_{\text{NONPASS}} \wedge \phi = 1\ (\small{\text{y}}).
\end{cases}
\end{align}
where $t_{tl}^{\text{next}}(t)$ is the remaining signal time on the current signal phase of the next traffic light.
\subsubsection{Optimal Control Problem}
We solve the following constrained finite-time optimal control problem for every $\mathcal{C}_{\text{pass}}$ candidates in the set \eqref{eq:C_pass} and collect the optimal cost $J^\star_{\text{ftocp}}(x(t),u(t),\mathcal{C}_{\text{pass}})$:
\begin{subequations}\label{opt:ocp}
\begin{align}
 \min_{\substack{\boldsymbol{x,u}}}&\; \sum\limits_{k=t}^{t+N-1}[\!\mathcal{E}(v_{k|t},u_{k|t})] + J^{LK}_{\text{smooth}}(\{x_{k|t}\}^{t+N-1}_{t}) \label{eq:cost} \\
 \text{s.t. }&\quad x_{k+1|t}=Ax_{k|t}+B u_{k|t},\label{eq:ev_dyn}\\
&\quad x_{k|t} \in \mathcal{X}, u_{k|t} \in \mathcal{U}, \\
&\quad (x_{k|t},u_{k|t}) \in \mathcal{F}^{LK}(s^{\text{front}}_{k|t},v^{\text{front}}_{k|t}), \label{eq:collision_avoidance}\\
&\quad x_{t|t}=x(t), ~\forall k\in\mathbb{I}_t^{t+N-1} \label{eq:initial_cond} \\
&\quad x_{k|t} \in \mathcal{P}(tl^{\text{next},s}), ~\forall k \in \mathbb{I}_{[t+\underline{t}_{\text{pass}}(\mathcal{C}_{\text{pass}},tl^{\text{next},p})]/\Delta t}^{[t+\bar{t}_{\text{pass}}(\mathcal{C}_{\text{pass}},tl^{\text{next},p})]/\Delta t}\label{eq:passing_constr}\\
&\quad x_{k|t} \in \mathcal{S}(tl^{\text{next},s}),~\forall k \in \mathbb{I}_{[t+\underline{t}_{\text{stop}}(\mathcal{C}_{\text{pass}},tl^{\text{next},p})]/\Delta t}^{[t+\bar{t}_{\text{stop}}(\mathcal{C}_{\text{pass}},tl^{\text{next},p})]/\Delta t}\label{eq:stopping_constr}
\end{align}
\end{subequations}
where $x_{k|t}$ denotes the decision variable at time step $k$ predicted at time $t$ and $N$ is the prediction horizon. In this work, we use $N=140$ to achieve 1 $Hz$ lane selection planning with the available hardware. The $\mathbb{I}_{k_1}^{k_2}$ notation denotes the set $\{k_1,k_1+1,\dots,k_2\}$. The optimal control problem has three components for the cost function. The first component is the energy minimization cost which uses the data-driven energy consumption model: $\mathcal{E}$, which is quadratic and convex w.r.t. the decision variables. The second component $J^{LK}_{\text{smooth}}(\{x_{k|t}\}^{t+N-1}_{t})$ is the stage cost added for smoothness of the
longitudinal trajectory and passenger comfort \cite{arpae-lc}.
After the optimal control problem is formulated in CasADi \cite{casadi} and solved using IPOPT \cite{ipopt}, the passing velocity of the first traffic light $v^\star_{\text{pass}}$ is extracted from the optimal solution, which is used to initialize the graph \eqref{eq:graph}. After computing \eqref{eq:J_graph}, we define
\begin{align}
    J^\star(&\mathcal{C}_{\text{pass}})= J_{\text{graph}}^\star(v^\star_{\text{pass}}) + J^\star_{\text{ftocp}}(x(t),u(t),\mathcal{C}_{\text{pass}}).
\end{align}
We compute the $J^\star(\mathcal{C})$ for every $\mathcal{C}_{\\text{pass}}\in\mathcal{LS}$. Finally, the optimization and graph search-based energy-efficient lane selector policy is
\begin{align}
\vspace{-0.1cm}
    \mathcal{C}^{\text{eco}}_{\text{pass}}=\argmin_{\mathcal{C}\in\mathcal{LS}} \; J^\star(\mathcal{C})
\end{align}
The lane selector also provides the reference velocity profile $v_{\text{ref}}$, which represents the energy-efficient speed trajectory corresponding to the selected behavior decision $\mathcal{C}^{\text{eco}}_{\text{pass}}$. The lower-level trajectory planner then computes a motion trajectory that closely tracks this reference velocity while respecting traffic laws and ensuring safety with respect to surrounding vehicles, using a finer time discretization to guarantee dynamic feasibility and reactivity \cite{arpae-lc}.
\section{Experiments}
\subsubsection{Experimental Setup} \label{sec:exp_setup}

\begin{figure}[h]
    \centering    \includegraphics[width=0.74\columnwidth]{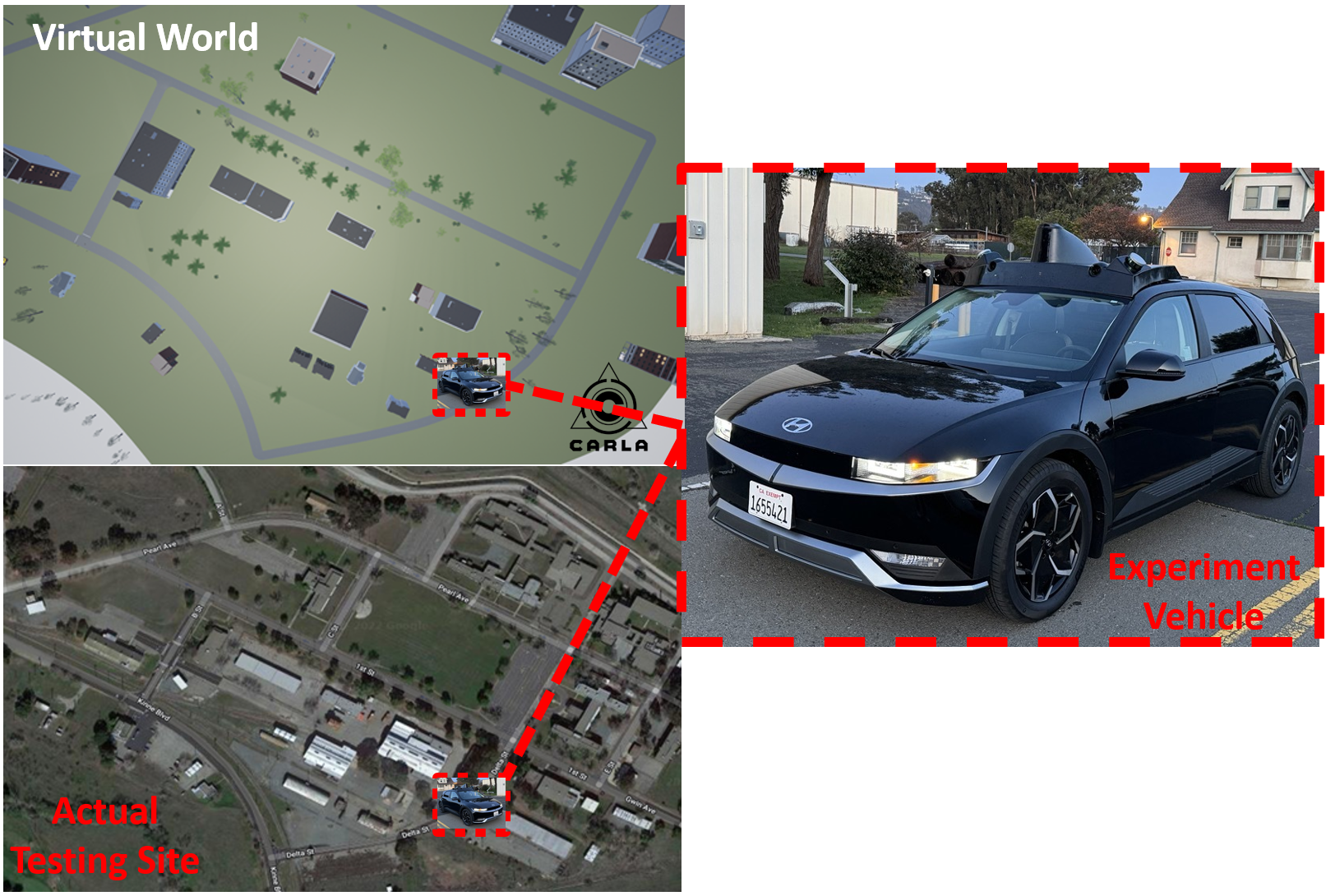}
    \caption{The virtual CARLA simulator map, the satellite image of the actual testing site, and the physical test vehicle}
    \label{fig:experiment_platform}
    \vspace{-0.2cm}
\end{figure}

We evaluate the proposed approach using a Vehicle-in-the-Loop (VIL) system that integrates a Hyundai IONIQ 5 with CARLA simulation for realistic traffic and SPaT signals. Fig. \ref{fig:experiment_platform} shows the virtual CARLA simulator map, the satellite image of the physical testing site, and the test vehicle (IONIQ 5). This setup enables robust evaluation of planning and control algorithms in a safe environment while using real energy measurements on the physical vehicle. See \cite{arpae-lc} for hardware details.

Each experiment consists of three laps around the route (approximately 4 km). For comparison, we designed two baseline approaches and conducted experiments to collect the corresponding energy consumption measurements: (i) a human driver observing the virtual environment (Baseline 1), and (ii) an autonomous eco-driving controller that optimizes longitudinal speed without lane changes (Baseline 2), which is essentially a lane-keeping algorithm that optimizes vehicle behavior based on an energy stage cost.


\subsubsection{Experiment Results} \label{sec:exp_result}
\begin{figure}[h]
\vspace{-0.2cm}
    \centering    \includegraphics[width=0.78\columnwidth]{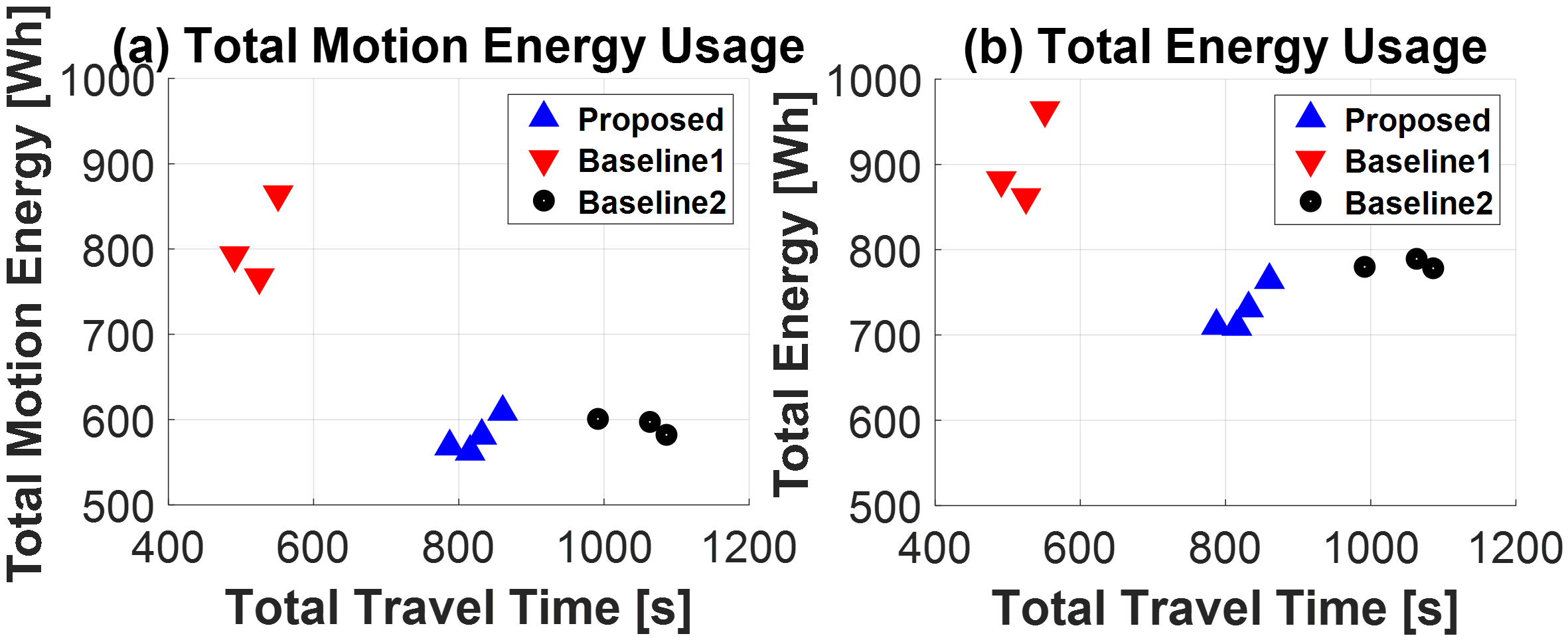}
    \caption{Experiment Results: (a) Total motion energy consumption (Overall energy consumption - auxiliary power usage); (b) Overall energy consumption.}
    \label{fig:exp_num_res}
\end{figure}

\begin{figure}[h]
    \centering    \includegraphics[width=0.72\columnwidth]{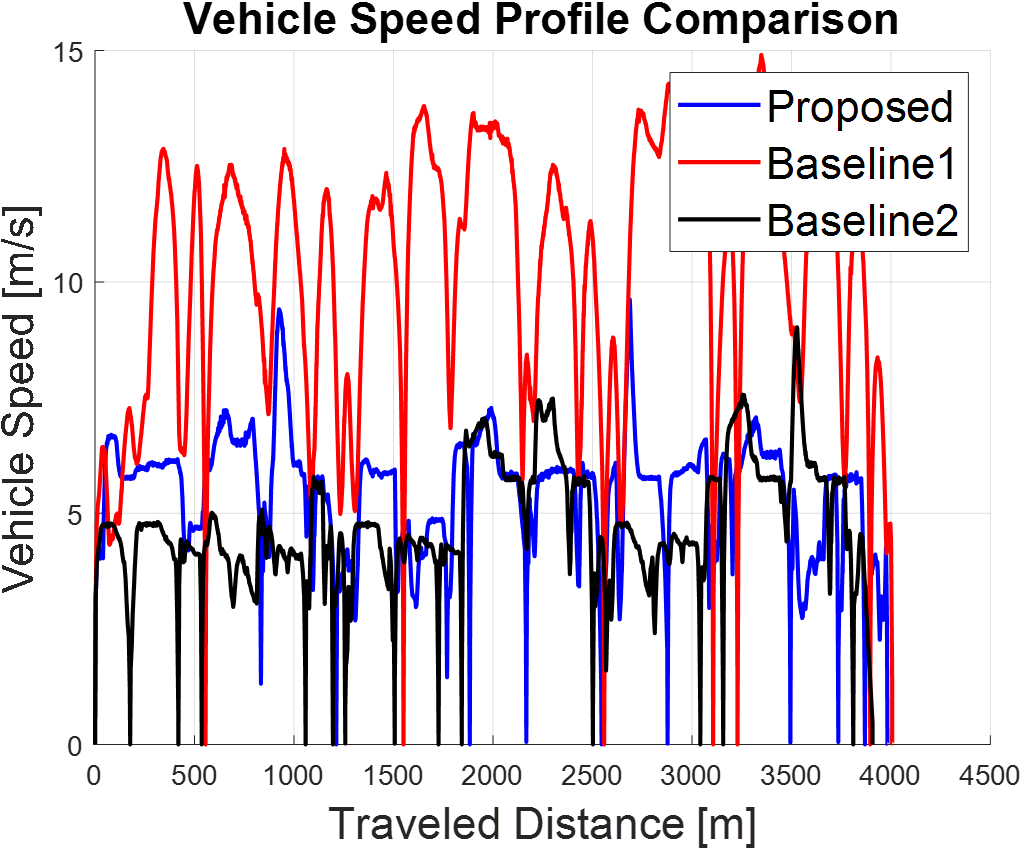}
    \caption{Experimental Results: Vehicle velocity profiles of a sample for each method in Fig. \ref{fig:exp_num_res}}
    \label{fig:exp_vel_plot}
\end{figure}

Fig. \ref{fig:exp_num_res}(a) presents the total motion energy consumption, calculated by subtracting auxiliary power usage from the overall energy consumption, while Fig. \ref{fig:exp_num_res}(b) shows the total energy usage including auxiliary loads. The human driver (Baseline 1) tended to drive closer to the speed limit and changed lanes as needed, resulting in the shortest average trip time of 522 seconds. In contrast, the proposed algorithm averaged 823 seconds per trip, and Baseline 2
resulted in the longest trip time of 1,047 seconds. These differences in longitudinal behavior are reflected in the velocity profiles shown in Fig. \ref{fig:exp_vel_plot}. 

As a result, the proposed algorithm achieved a 39\% reduction in motion energy consumption compared to Baseline~1, averaging $580$~Wh versus $808$~Wh. Furthermore, it achieved a 24\% reduction in total energy consumption (including auxiliary power usage) compared to Baseline~1 on average. While one might argue that the energy savings are simply due to the vehicle driving more slowly, the proposed algorithm actively computed energy-efficient velocity profiles and lane selections based on an energy cost function. No lower bound was imposed on the velocity profile, allowing the algorithm to explore the energy-optimal behavior of the electric vehicle.
Although Baseline~2 exhibited similar motion energy consumption to the proposed algorithm (within 2\% on average), its significantly longer trip time led to higher overall energy usage due to increased auxiliary power consumption, resulting in only a 7\% reduction on average. Ultimately, the proposed algorithm demonstrated the lowest total energy consumption as defined by the cost function. 
\vspace{-0.7cm}
\vspace{-0.5cm}
\section{Conclusion}

This work presented an energy-aware lane selection framework for 
connected electric vehicles that integrates lane-change decisions 
with longitudinal speed optimization using V2I communication. 
Vehicle-in-the-loop experiments with a Hyundai IONIQ5 demonstrated 
up to 39\% reduction in motion energy consumption compared to 
human driving and 7\% savings relative to a lane-keeping 
eco-driving baseline, without imposing explicit travel-time 
constraints.  

The results highlight the trade-off between energy efficiency and trip time, and show that lane selection is critical in reducing overall energy use for battery electric vehicles. While demonstrated on a specific route and traffic scenario, the framework illustrates how connectivity-enabled planning can improve the sustainability of autonomous urban driving.  

Future work will investigate the generalizability of the method across diverse environments and assess its robustness under varying traffic signal patterns and dynamic interactions with human drivers.
\section{Acknowledgments}
This research work presented herein is funded by the Advanced Research Projects Agency-Energy (ARPA-E), U.S. Department of Energy under DE-AR0000791. We also extend our sincere gratitude to the Hyundai Motor Company’s Vehicle Control Development Center and Autonomous Driving Development Center for their valuable technical assistance and support for the experimental vehicle.

\bibliographystyle{IEEEtran}
\bibliography{references}

\end{document}

%% file: macros.tex
\usepackage{verbatim}
\usepackage{graphics} 
\usepackage{amsmath,amssymb,mathtools,amsfonts}
\usepackage{multirow}
\usepackage{hhline}
\usepackage{algorithm}
\usepackage{footmisc}
\usepackage{flushend}
\usepackage[noend]{algpseudocode}
\usepackage{subcaption}
\usepackage{algorithmicx}
\usepackage{multirow}
\usepackage[utf8]{inputenc}
\usepackage{import}
\usepackage{graphicx, graphics}
\usepackage{caption}
\usepackage{subcaption}
\usepackage{array}
\usepackage{color}
\usepackage{siunitx}
\usepackage[short]{optidef}
\usepackage{float}
\usepackage{cite}
\PassOptionsToPackage{hyphens}{url}\usepackage{hyperref}
\usepackage{bbold}
\definecolor{myedit}{rgb}{1.0,0,0}
\definecolor{mytodo}{rgb}{0,0.0,1.0}
\definecolor{lightgray}{rgb}{0.8, 0.8, 0.8}

\DeclareMathOperator*{\argmin}{argmin}
